\documentclass[preprint,superscriptaddress,prl]{revtex4}
\usepackage{amssymb,amsmath,graphicx,amscd,xcolor}
\usepackage{float}
\begin{document}

\title{Large Zero Point Density Fluctuations in Fluids}

\author{Peter Wu}
\email{Peter.Wu610348@tufts.edu}
\affiliation{Institute of Cosmology, Department of Physics and Astronomy \\
Tufts University, Medford, Massachusetts 02155, USA}

\author{L. H. Ford}
\email{ford@cosmos.phy.tufts.edu}
\affiliation{Institute of Cosmology, Department of Physics and Astronomy \\
Tufts University, Medford, Massachusetts 02155, USA}

\begin{abstract}
Zero point density fluctuations in a liquid and their potential observation by light scattering are discussed. It is suggested that there
are two distinct effects of interest. One gives an average number of scattered photons, and depends upon an inverse power of the
photon wavelength. The second effect arises in the scattering of finite size photon wave packets and depends upon an inverse
power of the  spatial size of the wave packet. This effect appears as large fluctuations in the  number of scattered photons, and is
analogous to the vacuum fluctuations of spacetime averages of the energy density in quantum field theory.
\end{abstract}

\maketitle
\baselineskip=14pt	

Zero point fluctuations play an important role in several areas of physics, including both condensed matter physics and quantum
field theory. These fluctuations are a direct consequence of the uncertainty principle. When a quantum system is observed on
smaller length or time scales, the zero point fluctuations become greater. In quantum field theory, this means that the operators
describing observable quantities must be averaged over finite regions. The electric field or energy density at a single spacetime
point is not meaningful, but a space and time average is. This fact has a deep connection with quantum measurement and the
principle that only observable quantities are physically meaningful. Any measurement of a field operator necessarily involves an
average over finite space and time regions. 

The quantum fluctuations of a quadratic field operator, such as energy density, are especially subtle, with a probability distribution
which falls more slowly than exponentially~\cite{FFR12,FF15,FF20} . This leads to an enhanced probability for very large vacuum fluctuations for quantum
stress tensors, which can in turn drive large fluctuations of the gravitational field, a variety of quantum gravity effect~\cite{HV04}.

The zero point fluctuations in the density of a fluid provides an analog model for quantum stress tensor fluctuations, as well as being
an interesting phenomenon in its own right.   We can write the local mass density of the fluid as
\begin{equation}
    \hat{ \rho}(t,\mathbf{x}) = \rho_0 +\hat{ \rho_1}(t,\mathbf{x}) \,,
\end{equation}
where $\rho_0$ is the average mass density, and $\hat{ \rho_1}(t,\mathbf{x})$ is an operator which describes the local density fluctuations
around the mean value. This operator may be expanded in terms of phonon creation and annihilation operators as~\cite{LP69}
\begin{equation}
\hat{\rho_1}(t,\mathbf{x})=\sum_{\mathbf{q}}\sqrt{\frac{\rho_0 \hbar\omega}{2 V c_s^2}}\left[ {e}^{i(\mathbf{q}\cdot\mathbf{x}-\omega t)}b_{\mathbf{q}} 
+ {e}^{-i(\mathbf{q}\cdot\mathbf{x}-\omega t)}b^\dagger_{\mathbf{q}}\right]\,,
\label{eq:hatrho}
\end{equation}
where $V$ is a quantization volume and $\omega = c_s|\mathbf{q}|$, with $c_s$ the speed of sound in the fluid.
This operator is proportional to the time derivative of a massless scalar field~\cite{Unruh,FF04}, $\phi$:
\begin{equation}
\hat{\rho_1}(t,\mathbf{x}) = \frac{\sqrt{\rho_0}}{c_s^2} \, \dot{\phi}(t,\mathbf{x})\,.
\end{equation}
Here $\phi$ is identical to the relativistic massless scalar field, except with the speed of light replaced by the speed of sound,  $c\rightarrow c_s$.
Thus the fluctuations of $\hat{\rho_1}$ may be understood from those of $\dot{\phi}$.

The density fluctuations are potentially observable in light scattering experiments.
This effect was discussed in Ref.~\cite{FS09}, where the differential cross section for scattering by the zero point density
fluctuations was derived. Integrating over scattering angle, summing over final photon polarizations, and averaging on initial 
polarizations convert this result into a total cross section
\begin{equation}
\sigma_{0}=\frac{368\,\pi^4}{105}\;  \frac{\hbar \mathcal{V}\, \eta^4}{c_s\, \rho_0 \, \lambda_0^5} \,.
\label{eq:cross-section}
\end{equation}
Here  $\lambda_0$ is the vacuum wavelength of the light, $\eta$ is the index of refraction for the fluid, 
 and $\mathcal{V}$ is the effective scattering volume of the fluid.
As discussed in Ref.~\cite{FS09}, this effect may be viewed as Brillouin scattering by the density fluctuations, and the
$\lambda_0^{-5}$ dependence may be viewed as a product of a factor of  $\lambda_0^{-4}$, characteristic of 
Rayleigh-Brillouin scattering, and a factor of $\lambda_0^{-1} \propto \omega$ arising from the frequency spectrum of
zero point fluctuations. This cross section was derived assuming that the initial and final photon states are plane waves.

 In this letter, our primary interest will be a related, but distinct effect which depends crucially upon the initial photon state
 being a localized wave packet. We will examine a space and time average of the squared density operator, $\hat{\rho_1}^2$,
 or equivalently of $\dot{\phi}^2$. This averaging will be determined by the details of the wave packet and the scattering
 measurement, and will give an additional  contribution to the photon scattering. 
 Consider the average of the squared density operator over finite regions of space and time in the form
\begin{equation}
    \overline{:\hat{\rho_1}^2(t,\mathbf{x}):}=\int dt\: f(t) \int d^3x \:g(\mathbf{x}) \:\left(:\hat{\rho_1}^2(t,\mathbf{x}):\right)\,.
\label{eq:rho-sq}
\end{equation}
Here we have normal ordered this operator, as we are here concerned with its fluctuations around its mean value. 
The functions $f(t)$ and $g(\mathbf{x})$ are sampling function in time and space, respectively, and are assumed to have compact support,
meaning that they vanish outside of finite intervals. These functions will model the measurement process of the density fluctuations,
which is assumed to occur in a finite spacetime region. A class of such functions was constructed in Refs.~\cite{FF15,FF20}, and
may be characterized by their asymptotic forms of their Fourier transforms:
\begin{eqnarray}
\hat{f}(\omega)\sim e^{-|\omega\tau|^\alpha}, \:\:\: \omega \tau \gg 1,\:\:\: 0<\alpha<1
\label{eq:alpha}
\end{eqnarray}
and
\begin{eqnarray}
\hat{g}(\mathbf{k})\sim e^{-(k \ell)^\lambda}, \:\: k \ell \gg 1,\:\:\: 0<\lambda<1 \,,
\label{eq:lambda}
\end{eqnarray}
where we take the spatial function to be spherically symmetric for simplicity. Here $\tau$ and $\ell$ are the characteristic time and
space scales for the sampling, so $f(t)$ is nonzero over a time interval of order $\tau$, and similarly  $g(\mathbf{x})$ is nonzero in a sphere whose diameter is
of order $\ell$. The constants $\alpha$ and $\lambda$ determine the rates of switch-on and switch-off, with smaller values corresponding to more rapid switching.
The detailed relation between the value of  $\alpha$ and the  switch-on behavior of $f(t)$ was described in Sect.~IID of Ref.~\cite{FF15}.
A simple electrical circuit was also given in Ref.~\cite{FF15}  in which the current increases after a switch is closed in accordance with the $\alpha =1/2$, function,
so this is a physically realizable case. 

Let $\delta \rho^2 = \overline{(:\hat{\rho_1}^2:)}/\rho_0^2$  be the averaged fractional squared density fluctuations. The second moment, also the variance, of this operator
is, in the continuum limit,
\begin{equation}
\mu_2 =\langle(\delta \rho^2)^2\rangle =
\frac{\hbar^2}{128\pi^6 c_s^4 \rho_0}\int d^3q_1\:d^3q_2 \:\:\omega_1 \omega_2 \, \hat f^2(\omega_1+\omega_2)\,\hat g^2(\mathbf{q_1}+\mathbf{q_2})\,,
\end{equation}
which depends upon the details of the sampling functions. As discussed in Ref.~\cite{FF20}, time averaging is essential for the fluctuations to be finite.
If we were to let $f(t) =\delta(t)$, so that $\hat f(\omega) = 1$, then $\mu_2$ would diverge due to a large contribution from modes with 
$\mathbf{q_1} \approx -\mathbf{q_2}$.

Here we assume that the sampling functions arise from the shape of a probe wave packet of light, which
propagates through the fluid, and measures the density fluctuations by light scattering.   We take the packet to be approximately spherical with a size of
about $\ell$, so it takes a time of about $\tau \approx \eta \, \ell/c$ to travel a distance $\ell$, as the speed of light in the fluid is $c/\eta$. If we are able to detect
the photons scattered from the packet in this time, we have effectively measured $\delta \rho^2$. The root-mean-square value of a set of such measurements
is predicted to be $\delta\rho^2_{rms}=\sqrt{\mu_2}$. Because the speed of light is orders of magnitude larger than that of sound, we have 
$\ell=c\tau/\eta \gg c_s\tau$. In this case, we find
\begin{equation}
\mu_2 \approx \frac{(\hbar I)^2}{2^8\pi^4 \,\rho_0^2 \, \ell^3\, \tau^5\, c_s^7},
\label{eq:mu2}
\end{equation}
where $I^2$ is the dimensionless integral
\begin{equation}
I^2=\int_0^\infty dv \: v^2 \,\hat g^2(v/l) \int_0^\infty du \: u^4\, \hat f^2(u/\tau) \,.
\end{equation}
Now we can write
\begin{equation}
\delta \rho^2_{rms}  =  \sqrt{\mu_2} \approx \frac{\hbar I}{2^4\pi^2\, \rho_0 \, \ell^4}\sqrt{\frac{c^5}{\eta^5 \,c_s^7}}\,,
\label{eq:rms}
\end{equation}
where in the final expression we have set $\tau = \eta\, \ell/c$. A crucial feature of this result is that $\delta \rho^2_{rms} \propto \ell^{-4}$. This means that
the fluctuations of the squared density increase rapidly when probed on smaller scales. Note that the variance $\mu_2$ also depends upon the shape of the
wave packet through the dependence of $I^2$ on the parameters $\alpha$ and $\lambda$. In general, we can expect that smaller values of these parameters,
corresponding to more rapid switching and a slower rate of decrease of $\hat{f}$ and $\hat{g}$, will lead to a larger variance.
We can also calculate the third moment, $\mu_3$, to find that
\begin{equation}
\frac{(\mu_3)^{1/3}}{(\mu_2)^{1/2}}\propto \left(\frac{c_s \tau}{\ell}\right)^{1/2} \ll 1
\label{eq:skew}
\end{equation}
The above ratio  is small because $\ell \gg c_s\tau$, so the probability distribution for $\delta \rho^2$ is only very slightly skewed.

The factor of $1/\tau^5$ in Eq.~\eqref{eq:mu2} is a reflection of the need for time averaging, which was discussed above.
This factor in turn causes  $\delta \rho^2_{rms}$ in Eq.~\eqref{eq:rms} to be proportional  to the very large dimensionless ratio, $(c/c_s)^{5/2}$, 
which greatly enhances the magnitude of the squared density fluctuations in the fluid. Note that the density fluctuations we consider
are large because of this factor which enhances the variance. Here we do not address the asymptotic form of the probability distribution,
which depends upon higher moments. 

\begin{figure}[htbp]
\includegraphics[scale=0.3]{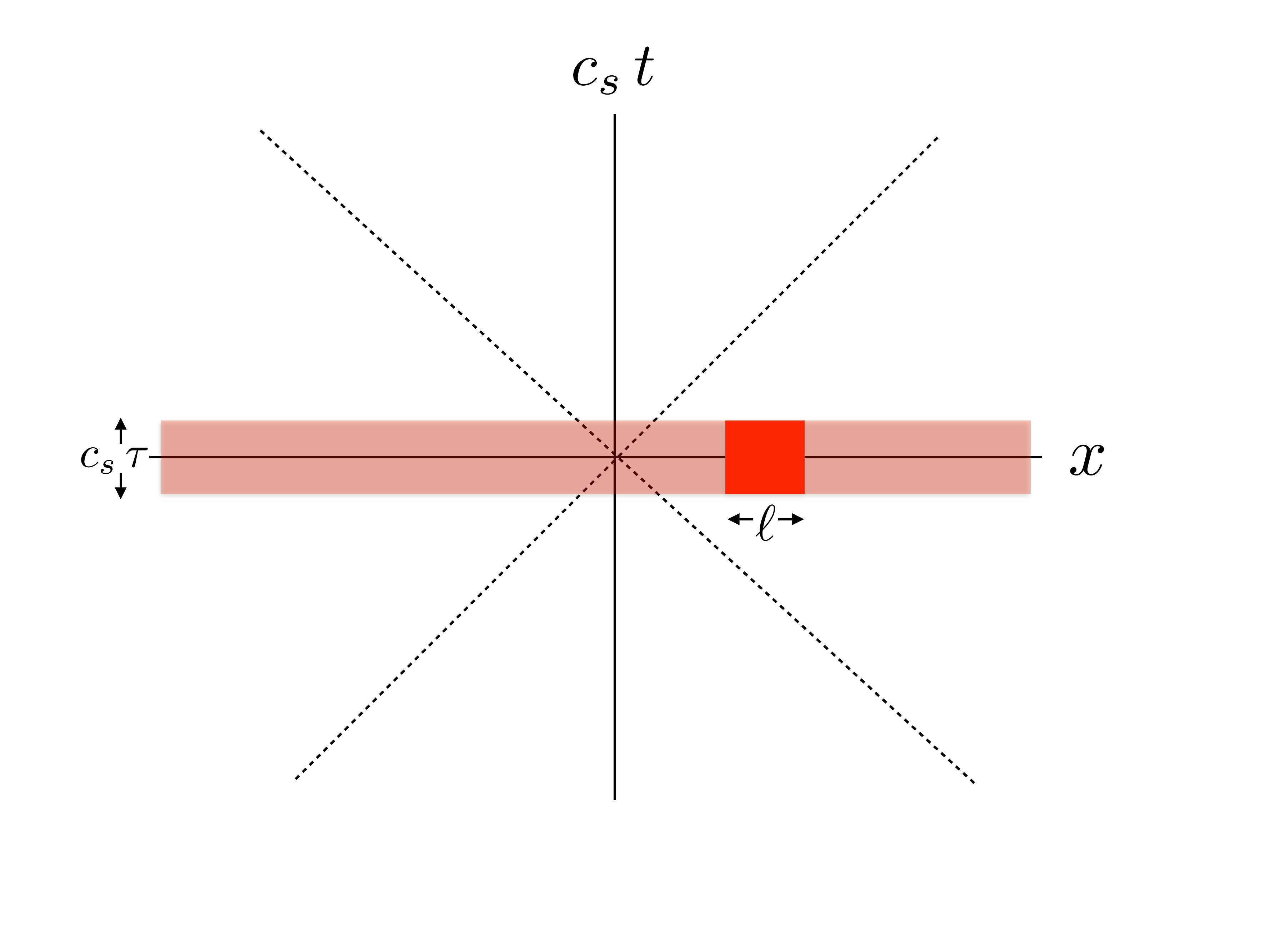}
\caption{A spacetime diagram for the fluid. The dotted lines are the soundcone, the spacetime paths of sound waves. The lightly shaded region 
is the spacetime path of a light pulse, which moves  in the $+x$-direction much more rapidly than sound. Here the pulse has a temporal 
duration of about $\tau$. The darker shaded region, whose spatial size is about $\ell$, represents the region which will scatter photons in
a particular measurement.}
\label{fig:soundcone}
\end{figure}

Let us consider a wave packet of light propagating through the fluid, as illustrated in Fig.~\ref{fig:soundcone}.
We take the spatial size of the packet in all directions to be of order $\ell$,
and its peak vacuum wavelength to be $\lambda_0 \ll \ell$. Here $\eta$ is the index of refraction of the fluid at the peak wavelength, and is 
assumed to be of order one. We wish to consider photons scattered from this packet in a time of $\tau = \eta \, \ell/c$, during which the packet
moves  distance of order $\ell$, so the scattering volume $\cal{V}$ will be of order $\ell^3$. The continuum approximation which we use seems
to require that both $\lambda_0$ and $c_s\, \tau$ be larger than the interatomic spacing. In particular, we take
\begin{equation}
c_s\, \tau \agt 10^{10}\,{\rm m}\,.
\label{eq:continuum}
\end{equation}

Let $n_\gamma$ be the approximate number of
photons in the packet, so the photon number flux is of order $n_\gamma/(\tau \ell^2)$. First consider the scattering effect described by
Eq.~\eqref{eq:cross-section}. The mean number of scattered photons in time $\tau$ is about
\begin{equation}
n_s \approx n_\gamma\frac{368(\pi n)^4\,\hbar \ell}{105\,c_s \,\rho_0\, \lambda_0^5}\,.
\label{eq:ns}
\end{equation} 
Because Eq.~\eqref{eq:cross-section} is a total cross section, $n_s$ includes photons scattered at all angles. The differential cross section
from which Eq.~\eqref{eq:cross-section} is derived (Eq.~(23) in Ref.~\cite{FS09}) is proportional to $\sqrt{1-\cos \theta}$, where $\theta$ is the
scattering angle. This means that back scattering, $\theta \approx \pi$, is somewhat more probable than scattering in other directions. 
Note that $n_s$ depends linearly upon $\ell$, but this is simply due to the increased scattering volume as $\ell$ increases.
In contrast,  $n_s$ increases as the the wavelength, $\lambda_0$, decreases.    We interpret $n_s$ as a mean 
number of scattered photons, and  $\delta \rho^2_{rms}$ as producing fluctuations around this mean value. 
Note that Eq.~\eqref{eq:cross-section} and hence Eq.~\eqref{eq:ns} strictly assume an initial plane wave state with wavelength $ \lambda_0$.
However, if our wave packet has a spatial size $\ell \gg  \lambda_0$, then the bandwidth will be small, $\Delta \lambda \ll \lambda_0$, and 
Eq.~\eqref{eq:cross-section} is still applicable.

\begin{figure}[htbp]
\includegraphics[scale=0.3]{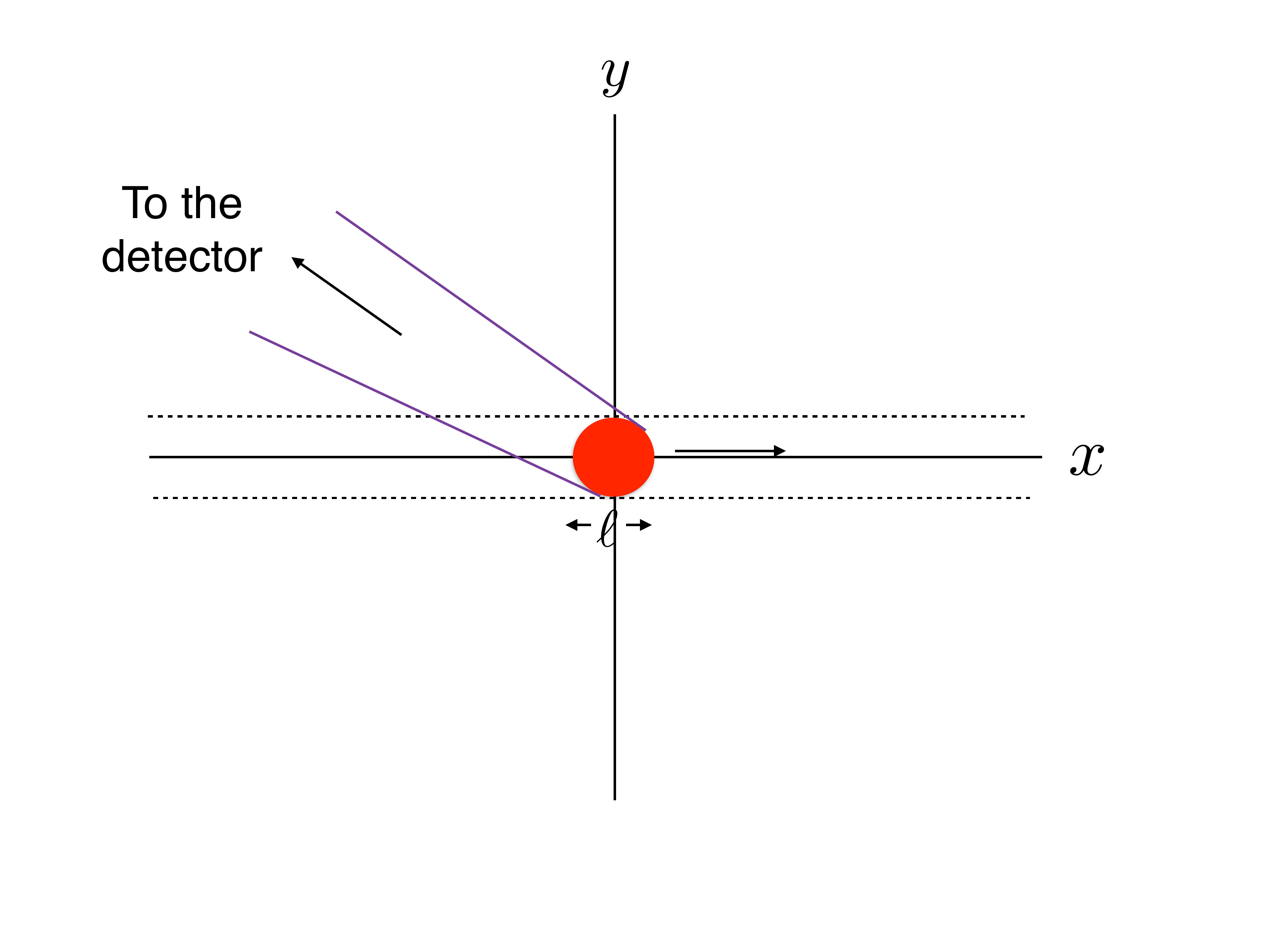}
\caption{The path of a wave packet of spatial size $\ell$ moving in  the $+x$-direction is illustrated by the dotted lines. The shaded region is the 
approximate position of the wave packet during a time interval of about  $\tau$, and is also the fluctuation region from which approximately 
backscattered photons will be detected. }
\label{fig:fluct-region}
\end{figure}

We may estimate the magnitude of the fluctuations due to $\delta \rho^2_{rms}$ by first considering an inhomogeneous medium with a local  variation in 
density on a length scale $\ell$, which scatters light with a much shorter wavelength. Model the boundary between
the region of varied density and the background medium as approximately a plane interface, and use the well-known results for Fresnel
scattering at such an interface. Let $\delta \eta$ be the fractional variation in index of refraction. Then
the reflection probability for normal incidence is about  $(\delta \eta)^2/4$ for both polarizations, and for wave packets as well as 
plane waves if the dispersion is small.  At other incidence angles, there is some 
dependence upon both the scattering angle and the polarization, but we will use the normal incidence results for an estimate. 
Note that if the magnitude of $\delta \eta$ is small, it is also approximately the fractional density variation. In our quantum treatment
of density fluctuations, the classical fractional density variation is replaced by $\hat{\rho_1}/\rho_0$, so $(\delta \eta)^2$ becomes $\delta \rho^2_{rms}$. 
 In the case of a pulse of light containing $n_\gamma$ photons, the expected variation in the number of photons scattered by a density fluctuation
 becomes
 \begin{equation}
\Delta n_s \approx \frac{1}{4}\, n_\gamma \,\delta\rho^2_{rms}.
\label{eq:Delta n_s}
\end{equation}
The scattering of a wave packet by a local density fluctuation is illustrated in Fig.~\ref{fig:fluct-region}.
In our view, the number of scattered photons, averaged over many trials, will be $n_s$, given by Eq.~\eqref{eq:ns}. However, in any one trial,
the number of scattered photons is likely to differ from $n_s$ by about $\Delta n_s$. This distribution will be skewed very slightly toward 
numbers larger than $n_s$, but smaller numbers of scattered photons are almost equally probable in light of Eq.~\eqref{eq:skew}, although
the associated probability distribution must vanish as the number of scattered photons goes to zero.    We can
view $\Delta n_s$ as arising from modifications of the phonon vacuum fluctuations due to the averaging produced by the finite wave packet
size and shape, and the choice to select photons scattered in a finite region, as illustrated in Fig.~\ref{fig:fluct-region}.
 
 We may combine Eqs.~\eqref{eq:rms}, \eqref{eq:ns} and \eqref{eq:Delta n_s} to obtain an expression for the expected fractional variation in scattered photon 
 number due to the modified vacuum fluctuations:
 \begin{equation}
\frac{\Delta n_s}{n_s} \approx \frac{105 \,I \, c^{5/2}\, \lambda_0^5}{23\cdot2^{10}\, \pi^6 \, c_s^{5/2}\, \eta^{13/2} \, \ell^5}\,.
\label{eq:Delta n_s-frac}
\end{equation}
 Before we discuss the possibility of observing this ratio, we need to consider statistical and thermal fluctuations. Let $\Delta n_{stat}$ be the expected
 statistical variation, so 
\begin{equation}
\frac{\Delta n_{stat}}{n_s} = \frac{1}{\sqrt{n_s}}  \approx\sqrt{\frac{105\,c_s \,\rho_0 \,\lambda_0^5}{ 368(\pi \eta)^4\, n_\gamma \hbar\, \ell}}\,,
\label{eq:Delta n_stat}
\end{equation}
which leads to
\begin{equation}
\frac{\Delta n_s}{\Delta n_{stat}} \approx \frac{I}{2^8\, \pi^4\, c_s^3}\,\sqrt{\frac{105\, n_\gamma\, \hbar\, c^5\, \lambda_0^5}{23\, \rho_0\, \ell^9\, \eta^9}}\,.
\label{eq:Delta n_stat2}
\end{equation}
The ratio between the differential cross sections of zero point and thermal fluctuations was found in Ref.~\cite{FS09} to be 
\begin{equation}
\frac{(d\sigma/d\Omega)_{ZP}}{(d\sigma/d\Omega)_{thermal}} \approx \sqrt{2(1-\cos\theta)}\frac{\hbar \pi c_s}{\lambda_0\, k_B \,T}\, \eta^4(\eta^2-1)^{-2}\,,
\end{equation}
where $T$ is temperature and $k_B$ is Boltzmann's constant. If we set $\theta = \pi$, for backscattering in the above expression, we can obtain an
estimate of the ratio of $n_s$ to $n_T$, the expected number of photons scattered by thermal density fluctuations:
\begin{equation}
\frac{n_s}{n_T} \approx \frac{2\hbar \pi c_s}{\lambda_0\, k_B \,T}\, \eta^4(\eta^2-1)^{-2}\,,
\label{eq:thermal}
\end{equation}
which implies
\begin{equation}
\frac{\Delta n_s}{n_T}\approx \frac{105\, \hbar\, c^{5/2}\, \lambda_0^4\,  I}{23 \cdot 2^9\, \pi^5\, c_s^{3/2}\, k_B\, T\, \eta^{5/2}\, (\eta^2-1)^2\, \ell^5} \,.
\end{equation}
The constant $I$ depends upon the choice of sampling functions. For a specific estimate, we set $\hat f(\omega) = \hat{h}_{fit}(\omega)$ and 
$\hat g (k)=\hat g_{fit}(k)$, where  $\hat{h}_{fit}(\omega)$ and $\hat g_{fit}(k)$ are defined in Appendix A of Ref~\cite{FF20}, and  correspond to
$\alpha = \lambda =1/2$ in Eqs.~\eqref{eq:alpha} and  \eqref{eq:lambda}. This choice gives $I\approx113$.

Now we wish to give some numerical estimates for a specific liquid, which we take to be He$^3$ in its normal (non-superfluid) phase.  
At $T = 1{\rm K}$ and atmospheric pressure,  $c_s \approx 200 {\rm m/s}$~\cite{Fairbank1966}, 
$\rho_0 \approx 83 {\rm kg/m}^3$~\cite{Edeskuty1960}, and  $\eta \approx 1.026$~\cite{Edwards1958,Chase1961}. Furthermore, in the temperature interval
$0.1{\rm K} \alt T \alt 1{\rm K}$, both $c_s$ and  $\rho_0$ were found in Ref.~\cite{AO1971} to be approximately independent of $T$. Given that the index of refraction is determined
by $\rho_0$ and the atomic polarizability through the Claussius-Mossotti relation, we can expect $\eta$ also to be  approximately independent of $T$.   
The lower bound on $\ell$ from the
validity of the continuum limit, Eq.~\eqref{eq:continuum}, is $\ell \agt 150 \mu {\rm m} $,  so let us consider a wavepacket of length $\ell \approx 400\mu {\rm m}$. 
For a wave packet of energy $E=1\mu {\rm J}$ peaked at a wavelength of 1 $\mu {\rm m}$, there are around  $n_\gamma = 5\times 10^{12}$ photons.
With this data, we find 
\begin{equation}
\frac{\Delta n_s}{n_s}\approx 0.12 \left(\frac{400\mu\text{m}}{\ell}\right)^5\left(\frac{1.026}{\eta}\right)^{13/2}\left(\frac{200 \text{m/s}}{c_s}\right)^{5/2}\left(\frac{\lambda_0}{1 \mu\text{m}}\right)^5,
\end{equation}
\begin{equation}
\frac{\Delta n_s}{\Delta n_{stat}} \approx 8 \left(\frac{400\mu \text{m}}{\ell}\right)^{9/2}\left(\frac{1.026}{\eta}\right)^{9/2}\left(\frac{200 \text{m/s}}{c_s}\right)^3
 \left(\frac{\lambda_0}{1 \mu\text{m}}\right)^{3}\left(\frac{E}{1\mu\text{J}}\right)^{1/2}\left(\frac{83 \text{kg/m}^3}{\rho_0}\right)^{1/2},
\end{equation}
and
 \begin{equation}
\frac{\Delta n_s}{n_T} \approx 4.5\left(\frac{400\mu\text{m}}{\ell}\right)^5\left(\frac{200 \text{m/s}}{c_s}\right)^{3/2}\left(\frac{\lambda_0}{1 \mu\text{m}}\right)^4 \left(\frac{0.1 K}{T}\right)\,,
\end{equation}
with $\eta \approx 1.026$.
In this case, we find that both the statistical and thermal fluctuation effects for $T \alt 0.5{\rm K}$ can be subdominant. Furthermore,  
$\Delta n_s$ can be a reasonable fraction of $n_s$, so it
seems that the effects of space and time averaging might be observable.

In summary, we have argued that there are two contributions to light scattering by zero point density fluctuations in a liquid. One gives an average number of
scattered photons and depends upon an inverse power of the photon wavelength. The other describes large fluctuations around this average, and depends
upon an inverse power of the size of the scattering region, which is determined by the size and shape of the photon wave packets and our choice of which scattered 
photons to count in a given measurement. This effect is closely analogous to the quantum stress tensor fluctuations expected in relativistic quantum field theory. 
The observation of both effects would reveal subtle features of quantum fluctuations.

\begin{acknowledgments} 
We would like to thank Norman Birge, Chris Fewster, and Roger Tobin for helpful discussions. 
This work was supported in part  by the National Science Foundation under Grant PHY-1912545.
\end{acknowledgments}

\end{document}